\documentclass[conference]{IEEEtran}
\IEEEoverridecommandlockouts
\usepackage{cite}
\usepackage{amsmath}
\usepackage{amsmath,amssymb,amsfonts}
\usepackage{algorithmic}
\usepackage{algorithm}
\usepackage{lscape}
\usepackage{graphicx}
\usepackage{textcomp}
\usepackage{xcolor}
\usepackage{multirow}
\DeclareUnicodeCharacter{2212}{\ensuremath{-}}
\def\BibTeX{{\rm B\kern-.05em{\sc i\kern-.025em b}\kern-.08em
    T\kern-.1667em\lower.7ex\hbox{E}\kern-.125emX}}
\begin{document}

\title{L2R-CIPU: Efficient CNN Computation with Left-to-Right Composite Inner Product Units\\
\thanks{This research was supported by the Basic Science Research Program funded by the Ministry of Education through the National Research Foundation of Korea  $(NRF-2020R1I1A3063857)$. The EDA tool was supported by the IC Design Education Center (IDEC), Korea.}
}

\author{
\IEEEauthorblockN{Malik Zohaib Nisar}
\IEEEauthorblockA{\textit{Department of Computer Engineering} \\
\textit{Chosun University}\\
Gwangju, Rep. of Korea \\
zohaib@chosun.ac.kr}
\and
\IEEEauthorblockN{1\textsuperscript{st} Given Name Surname}
\IEEEauthorblockA{\textit{dept. name of organization (of Aff.)} \\
\textit{name of organization (of Aff.)}\\
City, Country \\
email address}
\and
\IEEEauthorblockN{1\textsuperscript{st} Given Name Surname}
\IEEEauthorblockA{\textit{dept. name of organization (of Aff.)} \\
\textit{name of organization (of Aff.)}\\
City, Country \\
email address}
}

\author{\IEEEauthorblockN{Malik Zohaib Nisar, Muhammad Sohail Ibrahim, Muhammad Usman and
Jeong-A Lee}
\IEEEauthorblockA{Department of Computer Engineering,
Chosun University,
Republic of Korea\\
Email: \{zohaib,msohail,usman,jalee\}@chosun.ac.kr 
}}

\maketitle

\begin{abstract}
This paper proposes a composite inner-product computation unit based on left-to-right (LR) arithmetic for the acceleration of convolution neural networks (CNN) on hardware. The efficacy of the proposed L2R-CIPU method has been shown on the VGG-16 network, and assessment is done on various performance metrics. The L2R-CIPU design achieves $1.06 \times$ to $6.22 \times$ greater performance, $4.8 \times$ to $15 \times$ more TOPS/W, and $4.51 \times$ to $53.45 \times$ higher TOPS/$mm^2$ than prior architectures.
\end{abstract}

%%----------------------------------------------------------------------------------%%
%% 									PAPER KEYWORDS
%%----------------------------------------------------------------------------------%%

%%----------------------------------------------------------------------------------%%
%% 								MAIN CONTENT OF THE PAPER
%%----------------------------------------------------------------------------------%%
% \vspace{-5}
\section{Introduction}
Existing bit-serial convolutional neural network (CNN) accelerators perform computation in traditional right-to-left manner in which each arithmetic unit waits for the completion of the preceding operation before starting its computation. This introduces idle time for subsequent units, creating a bottleneck, slowing down the overall processing speed of the system and limiting the overall performance and scalability of the architecture, particularly in tasks where high throughput and low latency are critical \cite{ibrahim2024echo}. To overcome these challenges, we propose a CNN accelerator that employs a unconventional left-to-right (LR) computation pattern \cite{usman2023low}, in which the computations are executed serially in a most significant digit first (MSDF) order. For convolution computation in CNNs, a high throughput inner-product unit based on LR arithmetic is presented to enhance performance. To demonstrate the effectiveness of the L2R-CIPU approach, we examine our strategy on the convolutional layers of the VGG-16 network. The design has been compared with the conventional bit-serial design \cite{sharify2018loom} in terms of latency, performance, area utilization, and power consumption. Furthermore, the overall performance, performance per watt and performance per area of the L2R-CIPU accelerator is compared with two state-of-the-art accelerators. 

\section{Materials and methods}
% \vspace{-5}
\subsection{LR Inner Product Algorithm}
\begin{figure}[!ht]
	\begin{center}
 		\includegraphics*[width=4cm]{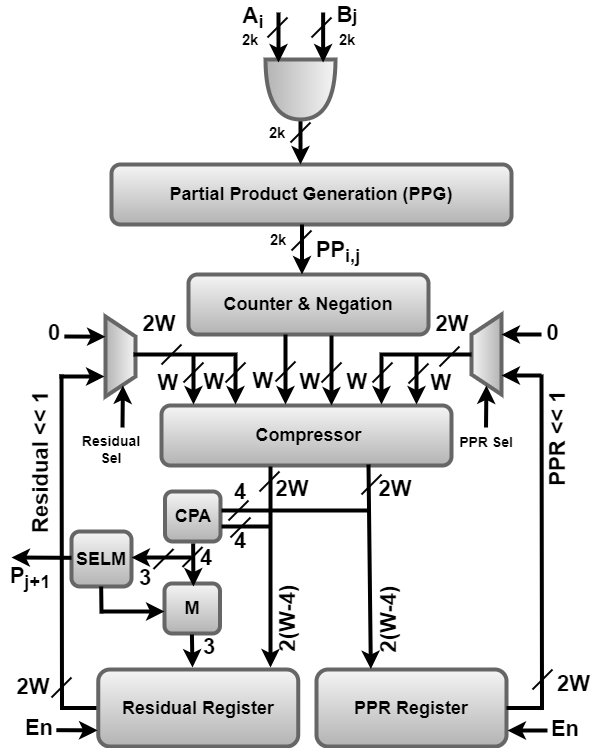}
 	\end{center}
 	\caption{LR Composite Inner Product Computation Unit \cite{2023hardware}}
 	\label{fig: OIP}
 \end{figure}

The LR arithmetic-based inner product unit is depicted in Fig.~\ref{fig: OIP}. The inner product unit simultaneously conducts $k$ discrete multiplications and accumulates their results in an online reduction tree, incurring a high cost of for the residual and PPR registers. All multiplications are merged into a single inner product unit such that a partial product term of $k$ multiplication $(A_{k,i}B_{k,j})$ is generated in each cycle which can be expressed as: $p = \sum_{k=1}^{k} A_{k} B_{k}$. By expanding  $A_{k}$ and $B_{k}$ we obtain; $ p = \sum_{k=1}^{k} \sum_{i=1}^{n} A_{k,i} 2^{-i} \sum_{j=1}^{n} B_{k,j} 2^{-j}$, which can be rearranged as: $p = \sum_{i=1}^{n} \sum_{j=1}^{n} \sum_{k=1}^{k} A_{k,i} B_{k,j} 2^{-(j+i)}$. These terms are then accumulated using a counter circuit.
\newline A row of partial product terms, are registered in \textit{Partial Product Row} (PPR), while residual is stored in \textit{residual register}. In each cycle, a row of the partial product array is generated by shifting the previous value stored in the PPR register and adding a new partial product term denoted as $PP_{i,j}$. A $6:2$ compressor performs the addition of values in residual register and PPR register with the new partial product term. Both these registers necessitate a bit width of $2\times$ the width of input operands, since they store values in carry-save format to prevent carry propagation during addition. Given that the residual register is updated every $n$ cycle, its value should not be added during those cycles. Hence, a multiplexer is integrated into the path of the residual value.
Similarly, the PPR register requires resetting every $n$ cycle. To achieve this, a multiplexer is utilized to input a zero value instead of the previous PPR value. Moreover, both the residual and PPR registers feature enabled signals to ensure that the output of the $6:2$ compressor is loaded only at the appropriate time.
% \vspace{-15}
\vspace{-0.5em}
\subsection{L2R-CIPU Design}
\begin{figure}[!ht]
	\begin{center}
 		\includegraphics*[width=0.65\linewidth]{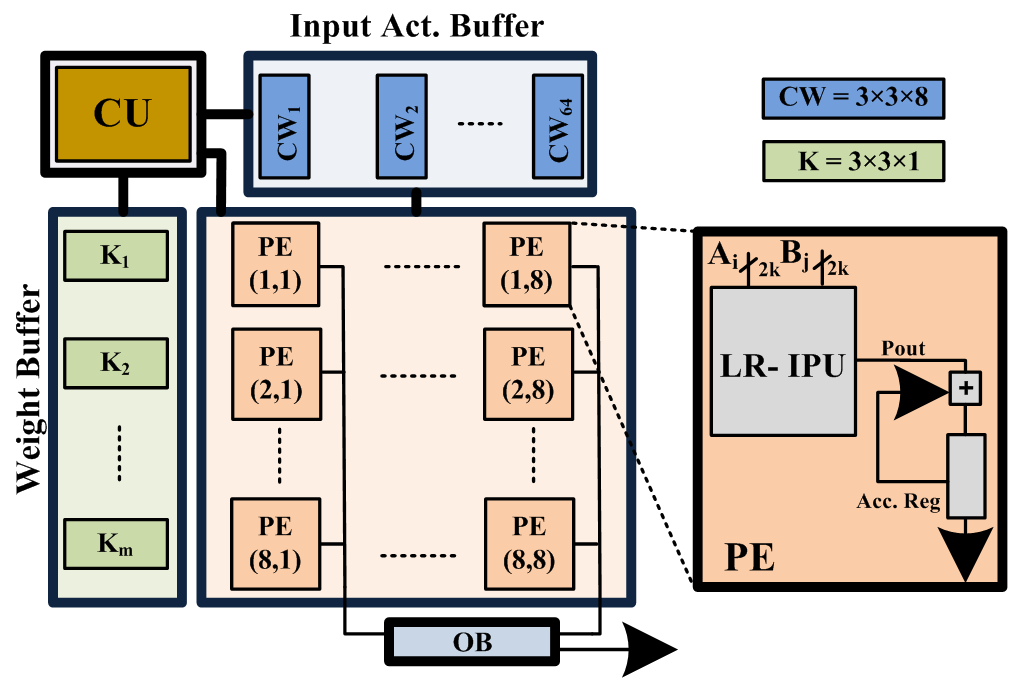}
 	\end{center}
 	\caption{Tiling and Processing Element of the L2R-CIPU Design }
 	\label{fig: OD}
 \end{figure}

We utilize the proposed inner product unit to form a processing element (PE). The PEs which serve as the fundamental units for convolution computations are arranged in a 2D array to form an accelerator as shown in Fig~\ref{fig: OD}. The configuration of the architecture is detailed later. Apart from PEs, the accelerator design encompasses several pivotal components, including control unit (CU), input activation/kernel buffers, and interconnects. The input activation buffers consist of convolution window (CW) buffers and the weight buffers consist of kernel buffers (K). The control unit generates signals to manage data, assisting in accelerator configuration which facilitates the communication between the activation and kernel buffer with the PEs. The network tile comprises an $8 \times 8$ PE array, tailored for processing of an input feature map window of size $3 \times 3$. The inner product unit within each PE is responsible for computing the sum of products (SOP) of a convolution window, across 8 input channels ($T_{n}=8$). After the convolution computation, the resulting partial product terms are accumulated to produce the final pixel, and the output is saved directly to the output buffer. The number of cycles for an inner product to generate its output is defined by  $\delta _{IP} = n^{2} + \delta_{Mult}$ cycles, where $n$ is the bit-width (input precision) and $\delta_{Mult}$ is the delay of the multiplier within the inner product unit. This delay is termed as online delay, after which the first most significant output digit is produced. The number of cycles for the accelerator is given by $Cycle_{P} = ((n^{2} + \delta_{Mult}) \times ((k \times k) + \lceil \frac{N}{T_{n}}\rceil) \times \lceil \frac{R \times C}{T_{r} \times T_{c}}\rceil  \times \lceil \frac{M}{T_{m}} \rceil$), where $\delta_{Mult}$ represents the online inner product delay, and $(k \times k)$ indicates the reduction stages in the adder tree required to generate the SOP for the $k \times k$ multipliers.$\frac{N}{T_{n}}$ represent the input feature map with an input tiling of $8$. Additionally, $\frac{R \times C}{T_{r} \times T_{c}}$ signifies the $8 \times 8$ tiling of rows and columns. $\frac{M}{T_{m}}$ denotes the output feature map with an output tiling of $1$.

\vspace{-1.5em}
% \vspace{-10}
\section{Result and Analysis}

The design is evaluated for several performance metrics including execution cycles, inference time, and GOPS performance for VGG-16 model. The RTL of the design is written using SystemVerilog and synthesis of both L2R-CIPU and the baseline architecture is carried out on Synopsys Design Compiler on NanGate $45$nm technology at a frequency of $400$ MHz.
% Figures \ref{fig: PP} and \ref{fig: Duration} depict the performance and duration of the convolution layers within the VGG-16 network, demonstrating significant improvements in performance and inference time compared to the baseline design. 
L2R-CIPU achieved remarkable performance gains of $3.40 \times$ for VGG-16 compared to our baseline design which uses computation pattern of \cite{sharify2018loom}. The comprehensive findings are summarized in Table~\ref{tab:syn}. To demonstrate the superiority of our design, we compared it with various existing CNN accelerators, as outlined in Table~\ref{tab:comparison}.  Our approach showcases significant advantages regarding high performance, rapid response time, and energy efficiency. In comparison to \cite{cheng2024leveraging}, the L2R-CIPU design achieves performance and energy gains of $6.22\times$ and $15\times$, respectively. Furthermore, compared to Eyeriss \cite{chen2016eyeriss}, the L2R-CIPU design delivers significantly faster inference times and achieves $1.06\times$ and $6.31\times$ better performance and energy efficiency, respectively. The L2R-CIPU design improves the area efficiency, reaching approximately $4.51\times$ to $53.45\times$ more TOPS/$mm^{2}$. 
%Thus, integrating LR modules enhances performance, energy, and area efficiency while reducing latency.

\vspace{-1em}
\begin{table}[!ht]
\centering
\renewcommand{\arraystretch}{1.2}
\caption{Synthesis results of the L2R-CIPU accelerator and compare with the baseline approaches using GSCL 45nm technology}\label{tab:syn}
% \begin{center}

\resizebox{0.62\linewidth}{!}
{
\begin{tabular}{l|c|c} \hline \hline
\text{Parameter} & \text{Baseline\cite{sharify2018loom}}  & \text{L2R-CIPU
} \\ \hline \hline

Latency (ns) 
& 3.23
& 0.34  \\ \hline

Area ($\mu m^2$) 
&324,379.52
&244,394.24
\\ \hline

Power ($mW$) 
& 55.61
& 40.67 \\ \hline \hline %15755.90,  648.38
\end{tabular}
}
\label{synth}
% \end{center}
\end{table}

\vspace{-1em}

\begin{table}[!ht]
\renewcommand{\arraystretch}{1.2}
\caption{ Overall performance comparison}
\centering
\resizebox{\linewidth}{!}{
\begin{tabular}{l|c|c|c|c}
\hline \hline
Designs & \cite{cheng2024leveraging} & \cite{chen2016eyeriss} & Baseline\cite{sharify2018loom} & L2R-CIPU \\
\hline \hline
Technology (nm) & 40 & 65 & 45 & 45 \\ 
Frequency (MHz) & 500 & 200 & 400 & 400 \\
Precision (Bits) & 8 & 16 & 8 & 8 \\
Peak Performance (GOPs) & 7.87 & 46.04 & 14.40 & 48.97 \\
Total Inference Time (ms) & - & 4309 & 2.24 & 0.86 \\
Power (mW) & 91.84 & 236 & 55.61 & 40.67 \\
Peak Energy Efficiency (TOPS/W) & 0.08 & 0.19 & 0.25 & 1.20 \\
Peak Area Efficiency (TOPS/$mm^{2}$) & 19.19 & 3.75 & 44.40 & 200.45 \\
Network & LENET-5 & VGG-16 & VGG-16 & VGG-16 \\
\hline \hline
\end{tabular}
}
\label{tab:comparison}
\end{table}

\vspace{-0.8em}
\section{Conclusion}
% \vspace{-2}
A composite arithmetic unit for the computation of inner products based on LR arithmetic has been proposed. The  method shows enhanced performance in terms of area, power, latency and throughput, compared to the conventional arithmetic based CNN accelerators. To this end, up to $6.22\times$, $15\times$ and up to $53.45\times$ improvement in terms of overall performance, throughput per watt, and throughput per area respectively has been recorded for the accelerator compared to the existing hardware accelerators for CNN computation.

\bibliographystyle{IEEEtran}
%\bibliography{./bibliography/IEEEabrv,./bibliography/IEEEexample}
\bibliography{Reference.bib}
%%----------------------------------------------------------------------------------%%
\end{document}